\documentstyle[12pt,epsfig]{article}

\catcode`\@=11
\long\def\@makefntext#1{
\protect\noindent \hbox to 3.2pt {\hskip-.9pt  
$^{{\ninerm\@thefnmark}}$\hfil}#1\hfill}                

\def\@makefnmark{\hbox to 0pt{$^{\@thefnmark}$\hss}}  
        
\def\ps@myheadings{\let\@mkboth\@gobbletwo
\def\@oddhead{\hbox{}
\rightmark\hfil\ninerm\thepage}   
\def\@oddfoot{}\def\@evenhead{\ninerm\thepage\hfil
\leftmark\hbox{}}\def\@evenfoot{}
\def\sectionmark##1{}\def\subsectionmark##1{}}

\renewcommand{\thefootnote}{\fnsymbol{footnote}}

\newcounter{sectionc}\newcounter{subsectionc}\newcounter{subsubsectionc}
\renewcommand{\section}[1] {\vspace*{0.6cm}\addtocounter{sectionc}{1} 
\setcounter{subsectionc}{0}\setcounter{subsubsectionc}{0}\noindent 
        {\normalsize\bf\thesectionc. #1}\par\vspace*{0.4cm}}
\renewcommand{\subsection}[1] {\vspace*{0.6cm}\addtocounter{subsectionc}{1} 
        \setcounter{subsubsectionc}{0}\noindent 
        {\normalsize\it\thesectionc.\thesubsectionc. #1}\par\vspace*{0.4cm}}
\renewcommand{\subsubsection}[1]
{\vspace*{0.6cm}\addtocounter{subsubsectionc}{1}
        \noindent {\normalsize\rm\thesectionc.\thesubsectionc.\thesubsubsectionc. 
        #1}\par\vspace*{0.4cm}}

\newcounter{appendixc}
\newcounter{subappendixc}[appendixc]
\newcounter{subsubappendixc}[subappendixc]

\renewcommand{\appendix}[1] {\vspace*{0.6cm}
        \refstepcounter{appendixc}
        \setcounter{figure}{0}
        \setcounter{table}{0}
        \setcounter{equation}{0}
        \renewcommand{\thefigure}{\Alph{appendixc}.\arabic{figure}}
        \renewcommand{\thetable}{\Alph{appendixc}.\arabic{table}}
        \renewcommand{\theappendixc}{\Alph{appendixc}}
        \renewcommand{\theequation}{\Alph{appendixc}.\arabic{equation}}
        \noindent{\bf Appendix \theappendixc #1}\par\vspace*{0.4cm}}

\def\abstracts#1{{
        \centering{\begin{minipage}{12.2truecm}\footnotesize\baselineskip=12pt\noindent
        \centerline{\footnotesize ABSTRACT}\vspace*{0.3cm}
        \parindent=0pt #1
        \end{minipage}}\par}} 

\renewenvironment{thebibliography}[1]
        {\begin{list}{\arabic{enumi}.}
        {\usecounter{enumi}\setlength{\parsep}{0pt}
\setlength{\leftmargin 1.25cm}{\rightmargin 0pt}
         \setlength{\itemsep}{0pt} \settowidth
        {\labelwidth}{#1.}\sloppy}}{\end{list}}

\newcounter{itemlistc}
\newcounter{romanlistc}
\newcounter{alphlistc}
\newcounter{arabiclistc}

\newcommand{\fcaption}[1]{
        \refstepcounter{figure}
        \setbox\@tempboxa = \hbox{\footnotesize Fig.~\thefigure. #1}
        \ifdim \wd\@tempboxa > 6in
           {\begin{center}
        \parbox{6in}{\footnotesize\baselineskip=12pt Fig.~\thefigure. #1}
            \end{center}}
        \else
             {\begin{center}
             {\footnotesize Fig.~\thefigure. #1}
              \end{center}}
        \fi}

\newcommand{\tcaption}[1]{
        \refstepcounter{table}
        \setbox\@tempboxa = \hbox{\footnotesize Table~\thetable. #1}
        \ifdim \wd\@tempboxa > 6in
           {\begin{center}
        \parbox{6in}{\footnotesize\baselineskip=12pt Table~\thetable. #1}
            \end{center}}
        \else
             {\begin{center}
             {\footnotesize Table~\thetable. #1}
              \end{center}}
        \fi}

\def\@citex[#1]#2{\if@filesw\immediate\write\@auxout
        {\string\citation{#2}}\fi
\def\@citea{}\@cite{\@for\@citeb:=#2\do
        {\@citea\def\@citea{,}\@ifundefined
        {b@\@citeb}{{\bf ?}\@warning
        {Citation `\@citeb' on page \thepage \space undefined}}
        {\csname b@\@citeb\endcsname}}}{#1}}

\newif\if@cghi
\def\cite{\@cghitrue\@ifnextchar [{\@tempswatrue
        \@citex}{\@tempswafalse\@citex[]}}
\def\citelow{\@cghifalse\@ifnextchar [{\@tempswatrue
        \@citex}{\@tempswafalse\@citex[]}}
\def\@cite#1#2{{$\null^{#1}$\if@tempswa\typeout
        {IJCGA warning: optional citation argument 
        ignored: `#2'} \fi}}

 1
 1
 1

\font\ninerm=cmr9

\begin{document}

\centerline{\normalsize\bf THEORETICAL ASPECTS OF HEAVY-FLAVOUR}
\baselineskip=22pt
\centerline{\normalsize\bf ELECTROPRODUCTION}
\baselineskip=16pt

\centerline{\footnotesize J. SMITH \footnote{On leave from The Institute
for Theoretical Physics, SUNY at Stony Brook, NY 11794-3840, USA}}
\baselineskip=13pt
\centerline{\footnotesize\it Deutsches Electronen-Synchrotron DESY, 
Theory Group,}
\baselineskip=12pt
\centerline{\footnotesize\it Notkestrasse 85, D-22607 Hamburg, Germany}
\centerline{\footnotesize E-mail: jacksmit@mail.desy.de}

\vspace*{0.6cm}
\abstracts{We discuss three theoretical schemes to describe charm 
quark electroproduction.}
 
\normalsize\baselineskip=15pt
\setcounter{footnote}{0}
\renewcommand{\thefootnote}{\alph{footnote}}
\vspace{0.2cm}
\centerline{\it To appear in Proc. of "New Trends in HERA Physics",
Ringberg, May 1997.}
\section{Introduction}

The study of charm electroproduction  
has become an important issue in the extraction of parton densities
in the proton. The reason is that the charm content 
$F_{2,c}(x,Q^2,m^2)$, where $m=m_{\rm charm}$, 
has grown from around one percent of the contribution to the
structure function $F_2(x,Q^2)$ 
in the $x$ and $Q^2$ region of the 
EMC experiment\cite{emc} to around twenty-five percent in the
$x$ and $Q^2$ region of the H1\cite{h1} and ZEUS\cite{zeus} 
experiments at HERA.
Therefore the analysis of parton densities in the proton
can no longer be done without considering c-quark production.

Let us begin with a brief review of some of the technical points in the
calculations of c-quark electroproduction in QCD.
We consider $F_{2,c}(x,Q^2,m^2)$ which 
is usually much larger than $F_{L,c}(x,Q^2,m^2)$.
We work in a three flavour number scheme (TFNS) where the u, d and s 
are light mass quarks ($n_f = 3$). 
The c $\bar{\rm c}$ pair is produced from the gluon by the
Bethe-Heitler process (photon-gluon fusion) in leading order (LO). 
In next-to-leading order (NLO) both the Bethe-Heitler and the 
Compton processes involve all light mass quarks (and antiquarks) 
and the gluon. A NLO calculation organizes the 
contributions from the various Feynman diagrams according to the 
following formula
\begin{eqnarray}
&& F_{2,c}(3,x,Q^2,m^2) =
x \int_x^{z_{\rm max}}\frac{dz}{z}\Biggl\{
\frac{1}{3} \sum\limits_{k=1}^{3} e_k^2 
\Biggl[
\Sigma\Big(3,\frac{x}{z},\mu^2\Big)  
L_{2,q}^{\rm S}\Big(3,z,\frac{Q^2}{m^2},\frac{m^2}{\mu^2}\Big)
\\
&& 
+ G\Big(3,\frac{x}{z},\mu^2\Big) 
L_{2,g}^{\rm S}\Big(3,z,\frac{Q^2}{m^2},\frac{m^2}{\mu^2}\Big)
 + \Delta\Big(3,\frac{x}{z},\mu^2\Big)
L_{2,q}^{\rm NS}\Big(3,z,\frac{Q^2}{m^2}, \frac{m^2}{\mu^2}\Big) \Biggr]
\Biggr\}
\nonumber\\ &&
+ x e_c^2 
\int_x^{z_{\rm max}} \frac{dz}{z}
\Biggl[ \Sigma\Big(3,\frac{x}{z},\mu^2\Big)  
H_{2,q}^{\rm S}\Big(3,z,\frac{Q^2}{m^2},\frac{m^2}{\mu^2}\Big)
+  G\Big(3,\frac{x}{z},\mu^2\Big)  
H_{2,g}^{\rm S}\Big(3,z,\frac{Q^2}{m^2},\frac{m^2}{\mu^2}\Big)
\Biggr] \,,\nonumber
\end{eqnarray}
where $e_c$ is the charge of the c-quark. 
The variable $z$ is the partonic longitudinal momentum fraction
and  $z_{\rm max} = Q^2/(Q^2 + 4 m^2)$.
The function $\Delta$ is the non-singlet (with respect to the flavour 
SU(3) group) combination of light-mass parton densities, at the scale $\mu$. 
The function $\Sigma$ is the singlet 
combination of these densities
while $G$ is the density for the gluon. Further
$L_{2,k}$ and $H_{2,k} ( k=q,g )$ are the singlet and non-singlet
heavy-quark coefficient functions at scale $\mu$. 
Since they contain factors of $\alpha_s$
we have explicitly indicated that they depend on $n_f=3$. 
The $L_{2,k}$ describe the reactions 
where the virtual photon couples to the light
quarks (u, d, and s) and the $H_{2,k}$ the reactions 
where it couples to the c-quark. 

The NLO contributions to Eq.(1) were originally calculated 
in\cite{lrsn} yielding single-particle inclusive
transverse momentum and rapidity distributions
of the c-quark\cite{lrsn1}. 
The functions $L_{2,k}$ and $H_{2,k}$ were only available
in the form of two-dimensional integrals.
To speed up the computation of Eq.(1) we made two-dimensional
grids of values for $L_{2,k}$ and $H_{2,k}$ 
together with an interpolation routine
in\cite{rsn}. The NLO calculation was repeated in a completely
exclusive fashion in\cite{hs1} so 
that one can plot all the distributions containing the c-quark,
the $\bar{\rm c}$ antiquark and the additional parton. This program 
incorporates a fragmentation function so that one can calculate
distributions for $D^{\ast \pm}$ mesons. 
These programs have been used by many 
authors\cite{grs},\cite{or},\cite{Vogt},\cite{several}
to discuss the sensitivity of the NLO results to changes in parton
densities, renormalization/factorization scales, and  
the mass of the c-quark. Finally the asymptotic
formulae for $L_{2,k}$ and $H_{2,k}$  in the limit $Q^2 \gg m^2$
were calculated in\cite{bmsmn}. The latter formulae enable one to
discuss the transition from the calculated c-quark rate in
NLO perturbation theory in the TFNS with $n_f=3$ 
to a four flavour number scheme (FFNS) with $n_f=4$ 
and a massless c-quark\cite{bmsn1},\cite{bmsn2}. 

Now we outline the scheme used for these NLO calculations. 
The renormalization of the virtual graphs 
with loops containing
light quarks was done in the $\overline{\rm MS}$-scheme 
while loops with heavy quarks were subtracted at zero external 
momentum. This scheme
was originally proposed in\cite{cwz} so that the light quark sector
remains the same as in a purely massless pQCD calculation. Hence
there is a matching condition at the scale $\mu =m$ for both $\alpha_s$
and for the parton densities. To explain this 
consider QCD with only massless quarks. Then the number of 
flavours enters
via the factors $n_f$ in the $\beta$-function and in the gluon splitting
function $P_{gg}$. We solve the equation
\begin{eqnarray}
\frac{\partial}{\partial \ln \mu^2} \Big(\frac{\alpha_s}{\pi}\Big)    
=
- \beta_0 \Big(\frac{\alpha_s}{\pi}\Big)^2 
- \beta_1 \Big(\frac{\alpha_s}{\pi}\Big)^3 - ... 
\end{eqnarray}
where
\begin{eqnarray}
\beta_0 = \frac{1}{4} \Big(\frac{11}{3} C_A - \frac{4}{3} T_f n_f \Big)
\quad \,, \quad
\beta_1 = \frac{1}{16} \Big( \frac{34}{3} C_A^2 - 4C_F T_f n_f - 
\frac{20}{3} C_A T_f n_f\Big) \,,
\end{eqnarray}
by introducing the $\Lambda_{\overline {\rm MS}}(n_f)$.
The solution of Eq.(2) is
\begin{eqnarray}
\frac{\alpha_s(n_f,\mu)}{\pi} 
= \frac{1}{\beta_0 \ln(\mu^2/\Lambda_{\overline{\rm
MS}}^2 )} - \frac{\beta_1}{\beta_0^3} 
\frac{\ln(\ln(\mu^2/\Lambda_{\overline{\rm MS}}^2))}
{\ln^2(\mu^2/\Lambda_{\overline {\rm MS}}^2)} + ..
\end{eqnarray}
A matching condition is therefore necessary to go from a TFNS with
$n_f=3$ to a FFNS with $n_f=4$. We have
to change the value of $\Lambda_{\rm \overline{\rm MS}}$ as we
cross the scale $\mu=m$ to keep the condition that
$\alpha_s(4,\mu=m) = \alpha_s(3,\mu=m)$.
However heavy particles cannot be ignored in QCD 
because they exist in vacuum fluctuations. 
It is well-known\cite{ac} that effects from very heavy
particles can be absorbed into an equivalent field theory 
at a smaller scale by an appropriate 
renormalization of the coupling constants and fields (decoupling of 
heavy flavours). QCD is an unbroken gauge theory 
which does not have heavy particle decoupling
when renormalized in the $\overline{\rm MS}$-scheme as all
quarks have $m=0$. 
However the top quark cannot be considered massless and 
virtual top-quark loops yield explicit terms in $\ln \mu^2/m^2$ (here 
$m = m_{\rm top}$) so there has
to be a matching condition between say a 
five flavour theory and a six flavour theory. 
This two-loop matching condition on $\alpha_s$
was first worked out in\cite{bw} and corrected in\cite{lrv}. It reads
\begin{eqnarray}
&&\frac{\alpha_s(6,\mu)}{\pi} =
\frac{\alpha_s(5,\mu)}{\pi} +
\Big(\frac{\alpha_s(5,\mu)}{\pi}\Big)^2  \frac{1}{3} T_f \ln 
\Big(\frac{\mu^2}{m^2}\Big)  
+ \Big(\frac{\alpha_s(5,\mu)}{\pi}\Big)^3 
\Big[ \frac{1}{9} T_f^2 \ln^2\Big(\frac{\mu^2}{m^2}\Big)
\nonumber \\ &&
+ \frac{1}{12} \Big( 5C_A T_f - 3C_F T_f\Big) \ln
\Big(\frac{\mu^2}{m^2}\Big) 
+ \frac{13}{48} T_f C_F - \frac{2}{9} T_f C_A \Big] 
+...\,.
\end{eqnarray}
Note that $\alpha_s(6,\mu=m)= \alpha_s(5,\mu=m)$ in order $\alpha_s$ 
but not in order $\alpha_s^2$. 

The parton densities must satisfy similar matching conditions to Eq. (5). 
In a NLO calculation the collinear singularities are 
regularized by $n$-dimensional regularization. We set the mass 
factorization scale equal to the renormalization scale $\mu$ for simplicity.
Only the gluon-gluon splitting function 
\begin{eqnarray}
P_{gg}(n_f,z) = 2C_A \Big[ z \Big(\frac{1}{1-z}\Big)_+
+ \frac{1-z}{z} + z(1-z) \Big] + \frac{1}{6}(11 C_A - 4 T_f n_f)
\delta(1-z)\,,
\end{eqnarray}
depends on the number of light quarks $n_f$. The c-quark is 
considered heavy with mass $m=m_{\rm charm}$ 
and we calculate the $L_{2,k}$ and $H_{2,k}$ in Eq. (1) in
the TFNS. There is no charm density at the scale $\mu = m$ but it does exist
for $\mu > m$. This means that the light flavour densities 
$f_k(x,\mu)$, $k=u,d,s,\bar u,\bar d,\bar s$ satisfy  
\begin{eqnarray}
f_k(4,x,\mu=m) = f_k(3,x,\mu=m)\,.
\end{eqnarray}
\begin{figure}
 \begin{center}
  {\unitlength1cm
   \epsfig{file=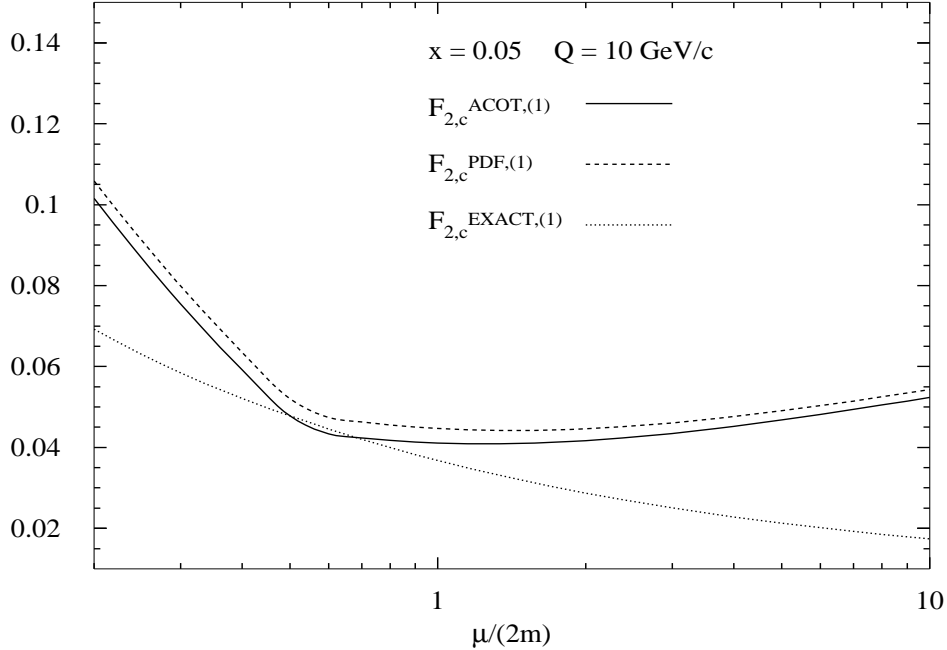,bbllx=520pt,bblly=95pt,bburx=105pt,bbury=710pt,%
      height=7.5cm,width=12cm,clip=,angle=270}
    }  
 \end{center}
\caption{Scale dependence of $F_{2,c}(x=0.05,Q=10 \,{\rm GeV}/c)$.}
\label{fig:1}
\end{figure}
One can examine
the DGLAP equations\cite{dglap} for the parton densities
when one chooses a scale with $\mu > m$. 
By taking the difference of the solutions with $n_f=4$ and $n_f=3$ 
and using Eq. (6) one finds that only the gluon density changes 
in order $\alpha_s$, namely
\begin{eqnarray}
f_g(4,x,\mu^2) = f_g(3,x,m^2)
\Big[ 1 - \frac{1}{3\pi} T_f \alpha_s(3,\mu) \ln \frac{\mu^2}{m^2}\,,
\Big]
\end{eqnarray}
and there is a c-quark density, proportional to $\alpha_s(3,\mu)
\ln (\mu^2/m^2)$, which grows at the expense of the gluon density.
The changes in the light flavour densities begin in order $\alpha_s^2$.
Hence at scales $\mu \gg m$ we can switch to a  
FFNS which contains a (massless) c-quark density and a different
momentum sum rule. Eq.(8) is a
matching condition on the gluon density between the TFNS and the FFNS
at the scale $\mu = m$.

A way to implement a smooth matching of $F_{2,c}(x,Q^2,m^2)$
from the TFNS in Eq. (1) to a charm density description 
in the FFNS was proposed in\cite{acot} and we call it the ACOT
scheme. Their scale choice, which we will use in the rest of this article, is
\begin{eqnarray}
\mu^2 \!\!\!\!\!\!&&
=\,\,\,m^2+ k Q^2 (1- m^2/Q^2)^n \quad \mbox{for} \quad Q^2 > m^2\,,
\nonumber\\
&& =\,\,\, m^2 \quad \mbox{for} \quad Q^2 \leq m^2\,,
\end{eqnarray}
with $k=0.5$, $n=2$ and $m = 1.5 \,({\rm GeV}/c^2)$.
They proposed that 
$F_{2,c}^{\rm ACOT,(1)}(x,Q^2,m^2)$ should consist of three terms.
The first one is the TFNS LO process in Eq.(1), 
which has the correct kinematics near threshold. 
It contains the convolution of the 
gluon density in the proton with the LO hard scattering cross section.
The third term is a FFNS charm density, i.e., the 
coefficient function in the proton structure function $F_2(x,Q^2)$ is 
simply a $\delta$-function, 
and this should be the best description at $\mu^2 \gg m^2$. 
Finally the second term is the product of two convolutions. Namely that  
of the gluon density in the proton with the massless c-density
in the gluon and with the LO hard scattering term (a $\delta$-function).
This second term is taken with a negative sign so that it cancels
the FFNS c-density term at small scales $\mu^2 \approx m^2$ 
while it tends to cancel the TFNS term when $\mu^2\gg  m^2$. 
The sum of these three terms is more stable under a variation of 
the scale $\mu$ than each term separately. 
In Fig.1 we show the scale dependence of $F_{2,c}(x,Q^2,m^2)$
for the TFNS LO term from Eq.(1) (labelled EXACT,(1)), 
the ACOT description (labelled ACOT,(1) and the
FFNS charm density (labelled PDF,(1)). 
We choose the same CTEQ parton densities\cite{cteq} as in\cite{acot}.
The monotonic decrease in
the EXACT result, which is due to the decrease in $\alpha_s$,
is modified in the other two approaches. 
In Fig.2 we plot the three results for $F_{2,c}(x=0.01, Q^2)$ 
versus $Q = \sqrt{Q^2}$. 
One sees that there
is a large difference between the EXACT result (TFNS)
and the PDF charm density result (FFNS) especially as $Q$ increases. 
However the charm density result cannot be used at small $Q$ where
it is negative. As expected the ACOT result interpolates between the EXACT 
result at small $Q$ and the charm density PDF 
result at large $Q$. 
\begin{figure}
 \begin{center}
  {\unitlength1cm
   \epsfig{file=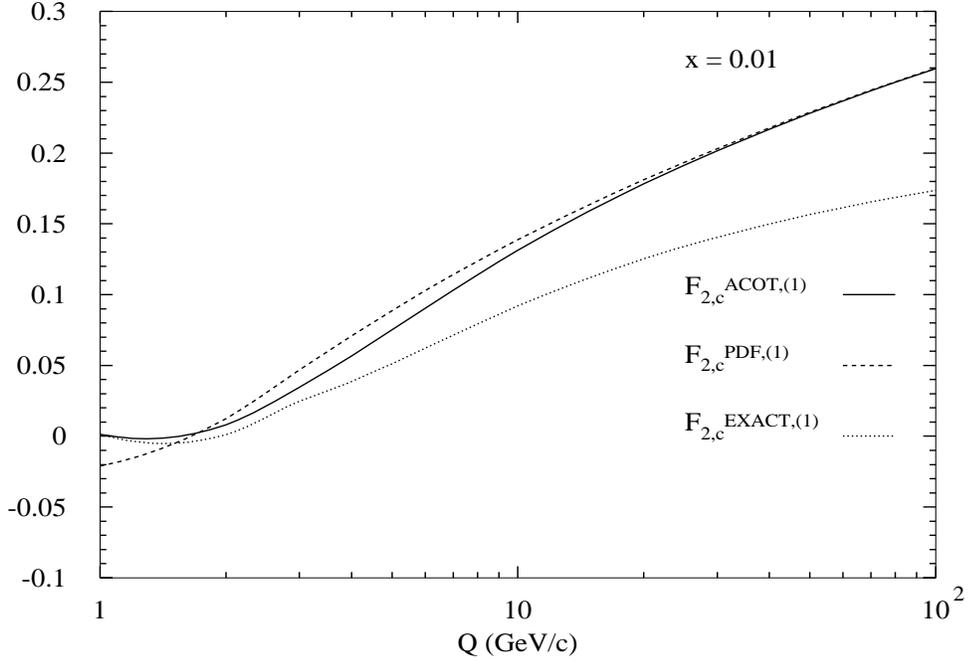,bbllx=520pt,bblly=95pt,bburx=105pt,bbury=710pt,%
      height=7.5cm,width=12cm,clip=,angle=270}
    }  
 \end{center}
\caption{$F_{2,c}(x=0.01)$ as a function of $Q$.}
\label{fig:2}
\end{figure}
The ACOT formula has been
incorporated into recent parton densities (the set
CTEQ4HQ\cite{lt}). Also there are now three light flavour (CTEQ4F3) and four
light flavour (CTEQ4F4) global fits to parton densities. The former is
basically equivalent to the parton density set in GRV94\cite{grv94}. An 
examination of the CTEQ4HQ set shows that the gluon is diminished from
previous sets\cite{cteq} and that all the light quark densities have changed,
not just the one for charm. This is due to the interplay between the gluon
and the light flavour densities in making a global fit to all the 
experimental data.
\begin{figure}
 \begin{center}
  {\unitlength1cm
   \epsfig{file=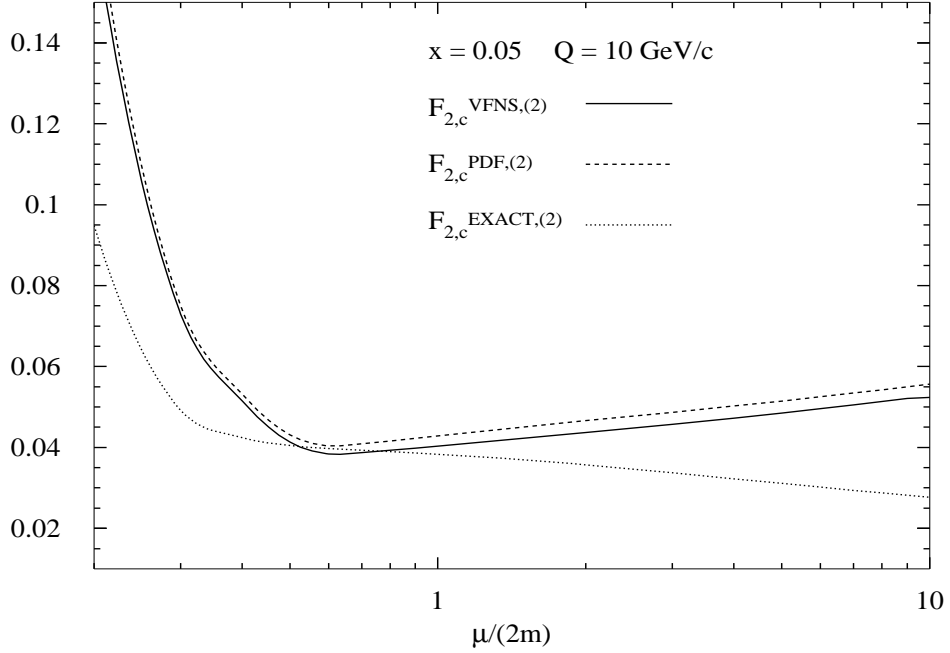,bbllx=520pt,bblly=95pt,bburx=105pt,bbury=710pt,%
      height=7.5cm,width=12cm,clip=,angle=270}
    }  
 \end{center}
\caption{Scale dependence of $F_{2,c}(x=0.05,Q=10 \,{\rm GeV}/c)$.}
\label{fig:3}
\end{figure}

As the ACOT description contains algebraic terms in $m$, it
is not obvious how to generalize it to higher orders in $\alpha_s$. 
(One proposal has been made in \cite{mrrs} which we cannot discuss here
for lack of space).
One would like to retain as much as possible a pQCD description based
on massless $\beta$ and $\gamma$ functions in the renormalization
group equation, and work in the $\overline{\rm MS}$ scheme. In\cite{bmsn1} 
we have worked out the two-loop matching conditions
on the flavour densities by the following procedure. Analytic
formulae for the functions $H_{2,k}$ 
in Eq.(1) are not known. In the 
limit that $Q^2 \gg m^2$ they reduce to expressions like 
$a(z)\ln^2 (Q^2/m^2) + b(z) \ln(Q^2/m^2) + c(z)$ (where we call them
$H^{\rm ASYMP}_{2,k}$). If we can find
$a(z), b(z)$ and $c(z)$
then we can mass factorize the logarithms into terms involving
$\ln(Q^2/\mu^2)$ and $\ln(\mu^2/m^2)$. If we then add the result 
to the two-loop light-flavour corrections\cite{zn1} to the proton
structure function $F_2(3,x,Q^2,\mu^2)$ then  
the terms in $\ln(Q^2/\mu^2)$ should recombine to yield the
two-loop corrections to $F_2(4,x,Q^2,\mu^2)$. 
Note that $F_{2,c}$ must be 
totally inclusive so we have to add the light quark vertex corrections
where the gluon propagator contains a charm quark loop,
which yields terms in $\delta(1-z)$. 
When the correct pieces have been identified we are left over with terms 
in $\ln^2(\mu^2/m^2)$ and $\ln(\mu^2/m^2)$ which cannot be
absorbed into the known matching conditions on $\alpha_s$. 
We can then define 
two-loop matching conditions for the parton densities to incorporate
these pieces, analogous to the two loop relations for
$\alpha_s$ in Eq. (5). 
We actually derived $L^{\rm ASYMP}_{2,k}$, $H^{\rm ASYMP}_{2,k}$ 
and the relations between the 
parton densities by calculating all two-loop matrix elements of the
operators in the OPE, where one loop contains 
a massless quark and the other loop a massive quark.  
\begin{figure}
 \begin{center}
  {\unitlength1cm
   \epsfig{file=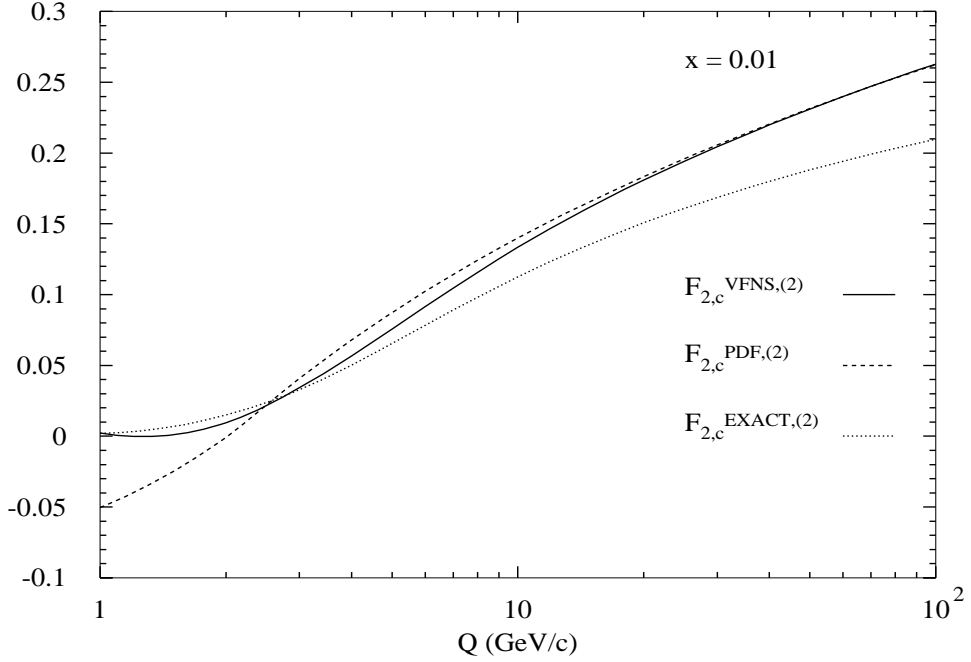,bbllx=520pt,bblly=95pt,bburx=105pt,bbury=710pt,%
      height=7.5cm,width=12cm,clip=,angle=270}
    }  
 \end{center}
\caption{$F_{2,c}(x=0.01)$ as functions of $Q$.} 
\label{fig:4}
\end{figure}

The result is that above the scale $\mu=m$ we have new FFNS parton densities
($n_f=4$) which are {\it calculable} from a TFNS set ($n_f=3$).
All densities are changed. The light quark densities pick up additional
terms in order $\alpha_s^2$, which are numerically rather small, but
are required for the FFNS parton densities to satisfy the momentum
sum rule.
The FFNS charm density is given in terms of the previous TFNS densities
and involves terms in order $\alpha_s$ and $\alpha_s^2$, i.e.,
\begin{eqnarray}
\lefteqn{f^{\rm FFNS}_{c + \bar c}(4,x,\mu^2) \equiv f_4(4,x,\mu^2) + 
f_{\bar 4}(4,x,\mu^2)}
\nonumber\\
&&= \Big( \frac{\alpha_s(\mu^2)}{4\pi}\Big)^2
\int_x^1 \frac{dz}{z} \Sigma\Big(3, \frac{x}{z} ,\mu^2\Big)
\tilde A_{cq}^{\rm PS,(2)}\Big(z, \frac{\mu^2}{m_c^2} \Big)        
\nonumber\\ 
&&+ \int_x^1 \, \frac{dz}{z} G\Big(3, \frac{x}{z}, \mu^2\Big)
\Big[  \Big( \frac{\alpha_s(\mu^2)}{4\pi}\Big)
\tilde A_{cg}^{\rm S,(1)}\Big(z, \frac{\mu^2}{m_c^2} \Big)
 +   \Big( \frac{\alpha_s(\mu^2)}{4\pi}\Big)^2
\tilde A_{cg}^{\rm S,(2)}\Big(z, \frac{\mu^2}{m_c^2} \Big) \Big]
\,,
\end{eqnarray}
where
\begin{eqnarray}
&& \hskip-0.5cm \tilde A^{{\rm PS},(2)}_{cq}\Biggl(\frac{m^2}{\mu^2}\Biggr)=
C_FT_f\Biggl\{
\Biggl[-8(1+z)\ln z-\frac{16}{3z}-4
%
%
+ 4 z +\frac{16}{3}z^2\Biggr] \ln^2\frac{m^2}{\mu^2}
\nonumber \\ && 
+\Biggl[8(1+z)\ln^2z-\Biggl(8+40z+\frac{64}{3}z^2\Biggr)\ln z
%
%
-\frac{160}{9z}
+16-48z+\frac{448}{9}z^2\Biggr] \ln\frac{m^2}{\mu^2}
\nonumber \\ && 
+ (1+z)\Biggl[32{\rm S}_{1,2}(1-z)+16\ln z{\rm Li}_2(1-z)
-16\zeta(2)\ln z
%
-\frac{4}{3}\ln^3z\Biggr]
+\Biggl(\frac{32}{3z}+8
\nonumber \\ && 
-8z-\frac{32}{3}z^2\Biggr) {\rm Li}_2(1-z)
%
%
+ \Biggl( -\frac{32}{3 z}-8+8z+\frac{32}{3} z^2\Biggr)\zeta(2)
%
+\Biggl(2+10z+\frac{16}{3}z^2\Biggr)  \ln^2z
%
%
\nonumber \\ &&
-\Biggl(\frac{56}{3}+\frac{88}{3}z
+\frac{448}{9}z^2\Biggr)\ln z
%
%
-\frac{448}{27z} - \frac{4}{3}
-\frac{124}{3}z+\frac{1600}{27}z^2 \Biggr\}  \,.
\end{eqnarray}
The other contributions to Eq. (10) and the relations between 
the other parton densities are available in\cite{bmsn1}. Eq. (11)  
shows that the charm density has a discontinuity at the 
scale $\mu=m$ when one considers order $\alpha_s^2$ terms, 
just as in Eq. (5). 
%

\begin{figure}
 \begin{center}
  {\unitlength1cm
\epsfig{file=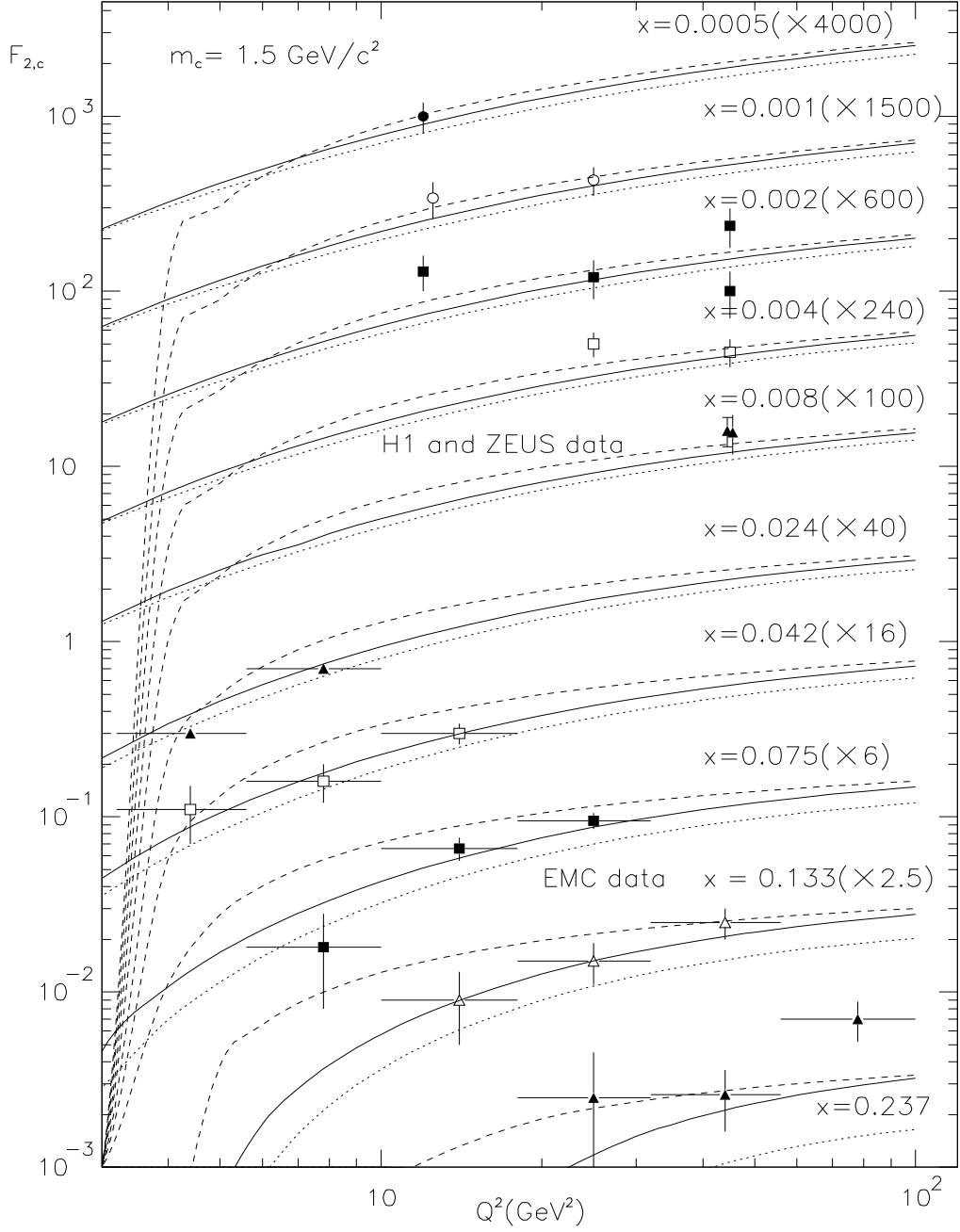,bbllx=50pt,bblly=100pt,bburx=510pt,bbury=720pt,height=18cm,width=14cm,clip=}
    }  
 \end{center}
\caption{
The functions $F_{2,c}^{\rm PDF,(2)}(4,x,Q^2)$ dashed line, 
$F_{2,c}^{\rm VFNS,(2)}(x,Q^2,m^2)$ solid line and 
$F_{2,c}^{\rm EXACT,(2)}(3,x,Q^2,m^2)$ dotted line, 
plotted as a functions of $Q^2$ for fixed $x$. The data points are from
the EMC, H1 and ZEUS experiments.}  
\label{fig:5}
\end{figure}

The above formula is for a FFNS renormalized in the
$\overline{\rm MS}$ prescription. 
Using the FFNS one can now calculate $F_{2,c}^{\rm PDF,(2)}(4,x,Q^2,m^2)$ 
in the region where $Q^2 \gg m^2$ and include the higher order QCD\cite{zn1}.
We expect this description to be valid in the region $Q^2 \gg m^2$ 
where the TFNS formula in Eq. (1) is  
dominated by the mass factorization logarithms. 
We have checked numerically that 
the asymptotic result for the NLO formulae, which we
call $F_{2,c}^{\rm ASYMP,(2)}(3,x,Q^2,m^2)$ is a good approximation to the
exact NLO result in Eq.(1), which we 
call $F_{2,c}^{\rm EXACT,(2)}(3,x,Q^2,m^2)$
when $Q^2\approx 10\, m^2$.
Hence above $Q^2 \approx 10\,m^2$ we have the option of retaining
the TFNS description for charm or to switching to the FFNS which sums large
logarithms in $\ln(Q^2/\mu^2)$ .

It now remains to find a formula which allows us to
interpolate between the 
TFNS and the FFNS and is valid in all orders in $\alpha_s$. 
In\cite{bmsn1} and\cite{bmsn2}
we have proposed the variable flavour number scheme (VFNS) formula 
\begin{eqnarray}
F_{2,c}^{\rm VFNS}(x,Q^2,m^2)
\!= \!F_{2,c}^{\rm EXACT}(3,x,Q^2,m^2)\! -\!F_{2,c}^{\rm ASYMP}(3,x,Q^2,m^2)
\!+ \!F_{2,c}^{\rm PDF}(4,x,Q^2)\,. 
\end{eqnarray}
This formula is designed like the ACOT one.
We have all the pieces to test it in order $\alpha_s^2$. (Full details 
are available in\cite{bmsn2}). The 
TFNS ASYMP term cancels the TFNS EXACT term at large $\mu$ 
leaving the FFNS charm density PDF term. 
However neither the ASYMP term nor the charm
density PDF term have the correct kinematics 
(they also have different $n_f$ and electric charges) 
to cancel each other exactly at small $\mu$ (near threshold). 
Using as input the same CTEQ densities\cite{cteq} as in\cite{acot},  
we have calculated (1) $F_{2,c}^{\rm EXACT,(2)}$,
the TFNS formulae in Eq.(1) 
(2) $F_{2,c}^{\rm VFNS,(2)}$, the VFNS formula in Eq.(12) and
(3) $F_{2,c}^{\rm PDF,(2)}$, the massless FFNS PDF result. 
In (3) we have evaluated the two convolutions to 
go from a TFNS to a FFNS and then to add the higher order 
corrections\cite{zn1}. In Figs. 3 and 4 we show
the analogous results to those in Figs. 1 and 2. One sees that the scale
sensitivity is mostly reduced for the TFNS EXACT result. The
PDF result is worse at small $Q$, which is a consequence of the fact that
some of the higher order contributions in QCD are negative. The VFNS
result Eq. (12) interpolates nicely between the other two.    
Finally in Fig. 5 we compare the three descriptions for $F_{2,c}$
with the data from the EMC, H1 and ZEUS experiments. Here we use the
GRV92\cite{grv92} parton densities.
One sees that the VFNS interpolation
is quite smooth as compared to that in\cite{mrrs}.
The massless FFNS results fall dramatically at small $Q$, 
due to the choice of $\mu$ in Eq. (9). 
All three curves are within the present experimental errors
of the H1 and ZEUS experiments. 
Therefore we will have to wait for more precise 
data before we can make a firm statement about which theoretical
approach is the best.

\section{Acknowledgements}
The work reported here was carried out in collaboration with 
M. Buza, Y. Matiounine and W.L. van Neerven.
The author thanks the
Alexander von Humbolt Stiftung for an award to allow him to spend his
Sabbatical leave in DESY.
%

\end{document}